\documentclass[aps,twocolumn]{revtex4}

\usepackage{amsmath,amssymb,hyperref}

\begin{document}

\title{Revisiting the Interpretations of Quantum Mechanics:\\

From FAPP Solutions to Contextual Ontologies.}

\author{Philippe Grangier}

\affiliation{Laboratoire Charles Fabry, IOGS, CNRS, 
Universit\'e Paris Saclay, F91127 Palaiseau, France.}

\date{\today}

\begin{abstract}
This note presents a concise and non-polemical comparison of several major
interpretations of quantum mechanics, with a particular emphasis on the
distinction between \emph{FAPP-solutions} (``For All Practical Purposes'') versus
\emph{ontological solutions} to the measurement problem.  Building on this
distinction, we argue that the Contexts--Systems--Modalities (CSM) framework,
supplemented by the operator-algebraic description of macroscopic contexts,
provides a conceptually complete, non-FAPP ontology that naturally incorporates
irreversibility and the physical structure of measurement devices.  This
approach differs significantly from other ontological interpretations such as
Bohmian mechanics, spontaneous collapse, or many-worlds, and 
highlights the major role of contextual quantization in shaping quantum theory.
\end{abstract}

\maketitle

\section{Introduction}

The literature on the interpretations of quantum mechanics is vast and recent
surveys \cite{list}  continue to map an increasingly diverse landscape of conceptual approaches.  Yet, despite this diversity, most interpretations fall into two
broad families: those that resolve the measurement problem only \emph{For All
Practical Purposes} (FAPP), and those that aim to provide a genuine
\emph{ontology} for quantum phenomena.  This distinction, introduced and
popularized by John Bell \cite{BellSpeakable}, remains central to contemporary debates.
\vskip 2mm

The purpose of this note is twofold.  First, it gives a brief, fair
comparison of several representative interpretations, emphasizing their
conceptual commitments rather than their technical details.  Second, it explains
how the Contexts--Systems--Modalities (CSM) framework 
\cite{CSM1,CSM2,Gleason,Uhlhorn}, especially when combined
with the operator-algebraic description of macroscopic contexts 
\cite{MP1,MP2,MP3}, occupies a
distinct position within this landscape: it is an ontological interpretation
that neither modifies the quantum formalism nor relies on FAPP arguments, and
it incorporates the real physics of macroscopic, irreversible measurements \cite{enigma}.
\vskip 2mm

It is nevertheless worth noting that CSM is often absent from
recent surveys, not because of conceptual incompatibility, but because it does
not fit neatly into the usual categories of hidden-variable, collapse, or
Everettian approaches. Therefore, completing the landscape 
motivates the present clarification.
\vspace{-3mm} 
\section{FAPP- and ontological solutions}
\vspace{-3mm} 
\subsection{FAPP-solutions.}
\vspace{-3mm} 
Bell introduced the acronym ``FAPP'' to describe a certain attitude toward the
measurement problem: the view that quantum mechanics works so well in practice
that conceptual difficulties can be set aside.  Decoherence theory has greatly
reinforced this attitude \cite{Zurek2003,darwinism}, by showing how interference between macroscopically distinct branches becomes negligible in realistic environments.  
Yet decoherence
does not explain why a single outcome is realized in an individual experiment;
it merely shows that the alternatives cease to interfere \cite{Zurek2003}, and that the measurement result  may be copied many times in the environment \cite{darwinism}.  For FAPP
interpretations, this is considered sufficient: the world looks classical 
in practice, and that is taken to be enough.
Therefore  the measurement problem is considered as effectively resolved once
decoherence, and possibly macroscopic mutiplication  are taken into account.  
Examples include:
\vskip 2mm

    -- Decoherence accounts of measurement \cite{Zurek2003,darwinism},
    
    -- Operationalist or Copenhagen-inspired interpretations (found in most textbooks, see e.g. 
    \cite{laloe,balian}),
    
    -- Subjectivist \cite{Zwirn2024} or QBist  \cite{Caves} approaches.
\vskip 2mm

Their strength lies in their pragmatic adequacy, which remains dominant in quantum
information science, where the focus is on operational tasks rather than
ontological commitments. Their weakness is that they
do not provide a fundamental ontology of outcomes: the measurement problem is
put aside rather than solved, and still open to discussion \cite{darwinism}. 

\vspace{-3mm} 
\subsection{Ontological solutions.}
\vspace{-3mm} 
Ontological interpretations take a different stance.  They insist that physics
must describe what exists, not merely what is observed, 
and must tell what are the physical properties of physical objects.   
They provide explicit entities: particles,
fields, collapses, branches, or other structures, and treat measurement as a
physical process governed by well-defined rules.
Examples include:
\vskip 2mm

 -- Bohmian mechanics (particle ontology,  \cite{Bohm1952}),

    -- Spontaneous collapse (GRW/CSL, \cite{GRW1986,Bassi2003}),
 
    -- Many-Worlds (branching ontology, \cite{Everett1957,deWitt}),
   
    -- CSM  (context-dependent ontology, \cite{CSM1,CSM2}).
\vskip 4mm

These approaches offer conceptual clarity but often require modifying the
formalism (GRW/CSL), adding hidden variables (Bohm), or adopting a global
ontology that is difficult to reconcile with everyday experience (Many-Worlds).
\vskip 2mm

The CSM framework  belongs to the ontological family, but with a distinctive twist: it introduces no hidden variables, no collapse mechanism, and no
branching worlds.  Instead, it proposes a contextual ontology in which physical
properties belong to quantum systems within  macroscopic measurement contexts.  This
move dissolves the measurement problem at its root, without altering the
quantum formalism. We will now give more details on this approach, 
and then come back to a comparison between  the different  options.

\section{The CSM framework}

The CSM framework introduces three primitive notions:
\begin{itemize}
    \item \textbf{Context}: the macroscopic measurement arrangement,
    \item \textbf{System}: the physical object under study,
    \item \textbf{Modality}: a set of certain, repeatable outcomes associated
    with a given system--context pair.
\end{itemize}

A modality is a well-defined, ontological property, but it is not a property of
the system alone: it is a property of the system within a context.  This
contextual ontology avoids the need for hidden variables, collapses, or
many  worlds.  Probabilities arise  when moving between incompatible
contexts, and Born's rule results from contextual quantization, that is 
the interplay between the quantized number of modalities that are accessible to a quantum system, and the continuum of contexts that are required to define these modalities; this is shown in details in \cite{Gleason,Uhlhorn}, based on Uhlhorn's and Gleason's theorems. .
\vskip 2mm

A central puzzle in quantum mechanics is that measurements yield single
outcomes, even though the quantum state may evolve into a superposition of
macroscopically distinct possibilities.  The usual presentation of the
measurement problem begins with a linear evolution that entangles system and
apparatus, producing a superposition of pointer states.  The question then
arises: why don't we observe superpositions of macroscopic results?
\vskip 2mm

In the CSM framework, this question does not show up, because the 
postulated ontology is based on the unicity of the macroscopic physical world \cite{MP3}, 
so that there is only one context at a time \cite{CSM1,CSM2}.
Measurement devices, detectors, screens, and magnets are all part of a single,
definite macroscopic universe.  They are not quantum systems that can be placed
in arbitrary superpositions of macroscopically distinct states.  
\vskip 2mm

A measurement
outcome is a modality: a definite, repeatable property of a system within
a given context.  Within each context the modalities are mutually exclusive and quantized; 
the outcomes are not superposed, they are selected by the context.
\vskip 2mm

This idea is reminiscent of Bohr's insistence on the
classical description of measurement devices, but CSM differs by providing a
precise ontology for the system-context pair rather than relying on 
some fluctuating ideas about complementarity. 
The statement ``there is no quantum world" attributed to Bohr
is replaced by ``classical contexts and quantum systems are both required
to build up the physical world". 
\vskip 2mm

This basic ontological choice has far-reaching consequences.  It means that the
``one result only'' fact is not something to be explained dynamically; it is a
structural feature of the ontology.  A modality is, by definition, a definite
property of a system in a definite macroscopic context.  There is no
``superposition of contexts,'' and therefore no ``superposition of modalities.''
A quantum superposition in the usual sense is a modality in another context. 
The measurement problem vanishes, because the physical ontology
itself excludes the problematic situation.
%
Once modalities are quantized, contextual, and mutually exclusive within one context, 
the appearance of
probabilities becomes unavoidable \cite{whatis}.  A modality defined in one context cannot
determine with certainty 
which modality will be realized when measuring in another, incompatible context.
Thus, a probabilistic description is required when moving between contexts.  The
Born rule emerges from the structure of these context changes, without the need
for collapse or branching \cite{Gleason, Uhlhorn}, clearing up many ``quantum mysteries". 
\vskip -4mm
\section{Infinite contexts and operator algebras}
\vskip -3mm
A distinctive feature of the CSM approach is the recognition that macroscopic
measurement contexts cannot be described mathematically as 
finite quantum systems.  Physically they involve extremely large numbers of
particles or degrees of freedom, and their evolution is generally 
dissipative, and irreversible.  Finite or countably infinite 
Hilbert spaces (technically, type I algebras) cannot capture these features, 
and to describe such contexts
mathematically, CSM employs the operator-algebraic framework of $C^*$ and 
von Neumann algebras \cite{Haag1996,BratteliRobinson}, as discussed in 
\cite{MP1,MP2,MP3}
\vskip 2mm

This move may appear unfamiliar to many physicists, partly because the
operator-algebraic formalism experienced a decline after its peak in the 1970s
and 1980s, and partly because it was often presented in a mathematically
abstract manner, without a clear physical motivation. Yet the mathematics itself
is well suited to the description of infinite tensor products of Hilbert spaces.  
What was missing was an ontology that explained what the algebraic structures represent
in an actual measurement.
\vskip 2mm

CSM provides this missing ontology.  A macroscopic context is represented by an
infinite algebra, often of type III, which naturally encodes
irreversibility and thermodynamic behavior.  A modality corresponds to a pure
state or sector within this algebra.  Changing context corresponds to moving
between inequivalent representations of the same algebra, reflecting the fact
that modalities are defined only within a given context.  Measurement is not a
unitary evolution on a single Hilbert space but a transition between
representations, consistent with the physical structure of macroscopic
apparatus.
\vskip 2mm

This reinterpretation rehabilitates operator algebras by giving them a clear
physical meaning: they provide the suitable mathematical tools to 
describe the classical behaviour of macroscopic contexts in which
modalities are defined.  In parallel, CSM provides the appropriate ontology
to  give a physical meaning to mathematical abstractions. 
We note also that 
this operator-algebraic treatment of contexts is absent from other ontological
interpretations.  Using it is a major asset, 
because it gives a natural explanation for the emergence of
irreversibility and the impossibility of superposing macroscopic contexts,
still keeping  the usual  quantum formalism to describe 
quantum systems within classical contexts.
\vskip 2mm

To avoid some misunderstandings, one may consider  
a usual spin measurement in a Stern-Gerlach (SG) experiment.
The SG magnet is part of the context, 
and its orientation decides which spin component is measured. 
But the active  type III
structure arises from the detectors placed in the output channels, 
which bring the result  to the macroscopic level. 
A simple model is obtained by coupling each output channel to an infinite tensor
product of two-level systems representing a spin chain or a field reservoir. The
detector begins in a thermal (KMS) state at finite temperature. The interaction Hamiltonian
induces a transition in one of the two reservoirs depending on the spin
projection. Because the reservoirs are infinite, the two resulting states belong
to inequivalent representations of the algebra of observables, and no unitary
transformation can connect them back after the measurement evolution. This inequivalence encodes irreversibility and the ``one result only'' structure.


\begin{table*}[ht!]
                            \centering
                            \renewcommand{\arraystretch}{1.8}
                            \setlength{\tabcolsep}{8pt}
                            \small                            
                           
              \begin{tabular}{|p{2cm}|p{1.7cm}|p{2.6cm}|p{1.8cm}|p{2cm}|p{1.3cm}|p{1.6cm}|}                           
                                          \hline
                                          \centering \textbf{Interpre-tation} &
                                          \centering \textbf{Ontology} &
                                          \centering \textbf{How outcomes become definite} &
                                          \centering \textbf{Role of context} &
                                          \centering \textbf{Macroscopic irreversibility} &
                                          \centering \textbf{Modifies QM predictions} &
                                          \centering \textbf{FAPP or ontological} \tabularnewline
                                          \hline
                                         
                                          \centering \textbf{Copenhagen, operational} &
                                          \centering None (instrumentalist) &
                                          \centering Postulated collapse &
                                          \centering External classical context &
                                          \centering Explained only FAPP &
                                          \centering No &
                                          \centering FAPP \tabularnewline
                                          \hline
                                         
                                          \centering \textbf{Decoherence-only} &
                                          \centering None, or wavefunction &
                                          \centering Interference suppressed by environment &
                                          \centering Emergent, not fundamental &
                                          \centering FAPP only &
                                          \centering No &
                                          \centering FAPP \tabularnewline
                                          \hline
                                         
                                          \centering \textbf{QBism / subjectivism} &
                                          \centering Agent-dependent beliefs &
                                          \centering Not addressed ontologically &
                                          \centering Context = agent’s action &
                                          \centering Not fundamental &
                                          \centering No &
                                          \centering FAPP \tabularnewline
                                          \hline
                                         
                                          \centering \textbf{Many-Worlds (Everett)} &
                                          \centering Universal wavefunction &
                                          \centering Branching of worlds &
                                          \centering Emergent from decoherence &
                                          \centering FAPP via decoherence &
                                          \centering No &
                                          \centering Ontological \tabularnewline
                                          \hline
                                         
                                          \centering \textbf{Bohmian mechanics} &
                                          \centering Particle positions + wave &
                                          \centering Particles have definite positions &
                                          \centering Not central &
                                          \centering Determinis- tic, reversible &
                                          \centering No &
                                          \centering Ontological \tabularnewline
                                          \hline
                                         
                                          \centering \textbf{GRW / CSL collapse models} &
                                          \centering Wavefunction + stochastic collapses &
                                          \centering Spontaneous collapse events &
                                          \centering Not central &
                                          \centering Built into dynamics &
                                          \centering Yes &
                                          \centering Ontological \tabularnewline
                                          \hline
                                         
                                          \centering \textbf{CSM with postulated contexts and systems} &
                                          \centering Modalities for system + context &
                                          \centering One modality realized per context;
                                          \centering probabilities from quantization&
                                          \centering Fundamental &
                                          \centering Not yet included explicitly &
                                          \centering No &
                                          \centering Ontological  \tabularnewline
                                          \hline
                                         
                                          \centering \textbf{CSM + operator algebras} &
                                          \centering Modalities for system + infinite context &
                                          \centering One modality realized per context;
                                          \centering probabilities from quantization &
                                          \centering Fundamental, operator algebras &
                                          \centering Encoded in type II/III algebras &
                                          \centering No&
                                          \centering Ontological, non-FAPP  \tabularnewline
                                          \hline
                            \end{tabular}
                            \caption{A comparative overview of several interpretations of quantum mechanics (QM) - see text}.
\end{table*}

\section{Comparison with other interpretations}

Now we come back to 
the broader landscape of interpretations, as represented in Table 1.
\vskip 2mm

On the FAPP side, 
Copenhagen and operationalist views rely on classical descriptions of
measurement devices and treat collapse as a pragmatic rule, 
or as a suitable approximation.  Decoherence-based
accounts explain the suppression of interference but do not fully address the
realization of single outcomes.  QBism and subjectivist approaches emphasize the
role of the agent and regard quantum states as expressions of belief rather than
physical properties. Realist-minded physicists may not feel fully happy there. 
\pagebreak 

For ontological interpretations, Bohmian mechanics \cite{Bohm1952} introduces particle
positions guided by the wavefunction, providing a clear account of outcomes but
at the price of nonlocal hidden variables.  GRW \cite{GRW1986} and CSL \cite{Bassi2003} 
modify the dynamics to
produce genuine collapses, introducing new parameters and stochastic processes.
They differ from other interpretations in the sense that their predictions differ from 
quantum mechanics; but the differences are extremely small, and have escaped 
any observation so far. 
Many-Worlds \cite{Everett1957} retains the linear dynamics but postulates a branching ontology in
which all outcomes occur, raising questions about the emergence of probabilities
and the status of branches. Though this approach is useful for quantum cosmology — it may be the reason for which it was promoted, in particular by DeWitt \cite{deWitt} — the world view  it provides is notoriously  counterintuitive. 
\vskip 1mm

We note that there are many other interpretations of quantum mechanics, not included in Table 1.  In particular we did not include Relational Quantum Mechanics (RQM)  \cite{Rovelli1996}, because it seems it is not currently in a fully stable state \cite{Calosi2024}. Also, it deals with an ontology of events (or ``flashes") rather than with an ontology of objects, and acknowledges ``universal unitarity". This  is not  compatible with CSM where unitary evolution is a feature of isolated subsystems. 
\vskip 1mm

CSM differs from all these approaches.  It retains the standard quantum
formalism, introduces no hidden variables or collapses, and does not rely on
branching.  Its ontology is contextual: physical properties are defined only
relative to macroscopic contexts.  When combined with operator algebras, this
ontology naturally incorporates irreversibility and the structure of
macroscopic measurement devices.  In this sense, CSM provides a non-FAPP
ontological interpretation that integrates the real physics of macroscopic
contexts. This solves the ``Measurement Problem" by grounding it in ontological realism,
telling that the measurement context and result belong to a unique macroscopic universe. 
This distinguishes CSM from the usual decoherence theory or RQM, who may be accused of being ``Everettian 
in denial", and also from the pure algebraic formalism, that stays in the realm of mathematical abstraction.
\vskip 1mm

It can be said that CSM is a neo-Bohrian or neo-Copenhagian interpretation, but these interpretations are on the FAPP side as explained above, whereas CSM in on the ontological side. One must emphasize nevertheless  that the CSM ontology is quite remote from the usual  reductionist ontology inherited from classical physics. Its philosophical  background  is rather contextual objectivity, as described in \cite{CO}. 
For a discussion shedding a new light on the historical Einstein-Bohr debate see \cite{BE}.

\vspace{-3mm}
\section{Conclusion}
\vspace{-3mm}
The distinction between FAPP-solutions and ontological solutions clarifies the
conceptual landscape of quantum interpretations.  Within this landscape, the
CSM framework occupies a unique position: it provides a non-FAPP ontology that
remains fully compatible with standard quantum mechanics, while incorporating
the real physics of macroscopic, irreversible contexts through the use of
operator algebras.
\vskip 1mm

On the practical side, an ontology anchored in 
contextual objectivity \cite{enigma,CO,BE} may also be
useful  for  building intuition for quantum technology engineers. 
\vskip 1mm

The CSM combination of ideas  offers therefore a new and coherent way to
understand quantum phenomena, one that complements---and in some respects
surpasses---existing ontological interpretations, by grounding contextuality and
irreversibility in the mathematical structure of infinite quantum systems.
\vskip 2mm

\textbf{Acknowledgements:} The author thanks Olivier Ezratty, Mathias Van Den Bossche, Roger Balian, Franck Lalo\"e and Herv\'e Zwirn for helpful discussions and comments. Copilot (online version  ``Smart", 2025) has been used for editing this paper.

\end{document}